\documentclass{moriond}
\usepackage[utf8]{inputenc}  
\usepackage[T1]{fontenc}
\usepackage{graphicx,psfrag}
\usepackage{mathrsfs}
\usepackage{amsmath,amsfonts,amssymb,pifont,gensymb}
\usepackage{multirow,enumerate}
\usepackage{comment,hyperref}
\usepackage{color}
\usepackage{acronym}
\usepackage{xspace}
\usepackage[normalem]{ulem}
\usepackage{mathtools}
\usepackage{subfigure}
\usepackage{makecell}
\usepackage{newtxtext,newtxmath}
\usepackage{changepage}
\usepackage{appendix}
\usepackage{lipsum}

\def\Journal#1#2#3#4{{#1} {\bf #2}, #3 (#4)}

\def\PRD{{\em Phys. Rev.} D}

\def\be{\begin{equation}}
\def\ee{\end{equation}}
\def\bea{\begin{eqnarray}}
\def\eea{\end{eqnarray}}

\newcommand{\boldtheta}{\boldsymbol{\theta}}

\newcommand{\Photo}{\includegraphics[height=35mm]{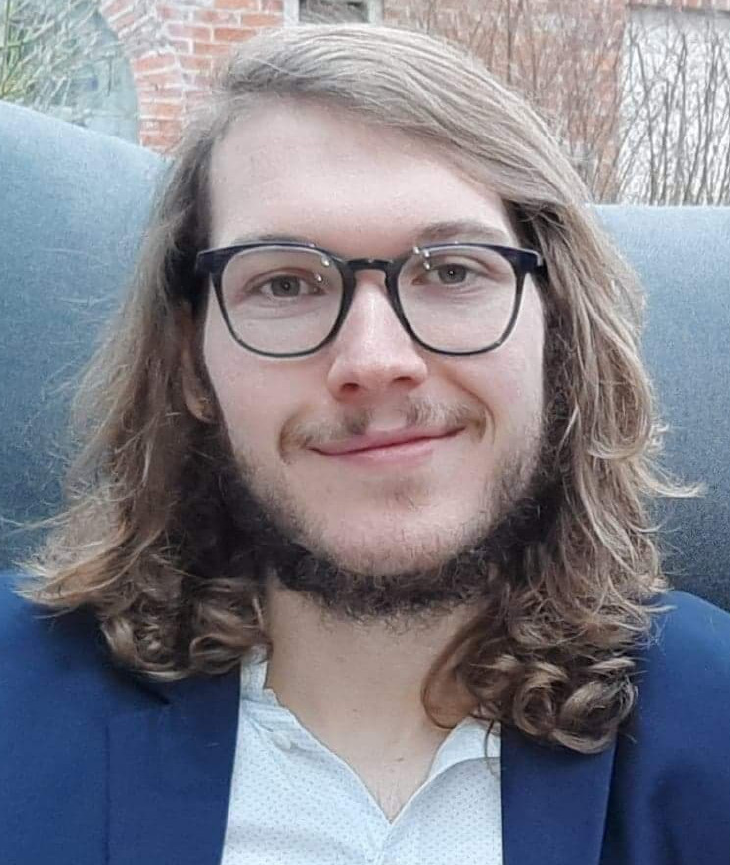}}

\begin{document}
\vspace*{4cm}
\title{GOLUM: A fast and precise methodology to search for, and analyze, strongly lensed gravitational-wave events}

\author{\textbf{J. Janquart}$^{1, 2}$, O.A. Hannuksela$^{3}$, K. Haris$^{1, 2}$, C. Van Den Broeck$^{1, 2}$}

\address{$^{1}$Nikhef – National Institute for Subatomic Physics, Science Park, 1098 XG Amsterdam, The Netherlands \\
$^{2}$Institute for Gravitational and Subatomic Physics (GRASP), Department of Physics, Utrecht University, Princetonplein 1, 3584 CC Utrecht, The Netherlands \\
$^{3}$ 
Department of Physics, The Chinese University of Hong Kong, Shatin, NT, Hong Kong}

\maketitle
\abstracts{Like light, gravitational waves can be gravitationally lensed by massive objects along their travel path. Strong lensing produces several images from the same binary coalescence and is forecasted to have a promising rate in ground-based gravitational detectors. To search for this effect in the data, one would, in principle, have to analyze all the possible combinations of the individual detected events, whose number will be ever-increasing. To keep up with the rising computational cost, we propose a fast and precise methodology to analyze strongly-lensed gravitational wave events. The method works by effectively using the posterior of the first image as prior for the second image. Thanks to its increased speed tractability, this method enables the joint analysis of more than two images. In addition, it opens the door to new strong lensing studies where a large number of injections is required.}

\section{Introduction}
When gravitational waves (GWs) travel from their source to the Earth, they can pass by massive objects (e.g. galaxies or galaxy clusters) and get gravitationally lensed~\cite{Takahashi:2003ix}. For strong lensing, the GW wavelength is larger than the typical size of the object, so that we are in the geometric-optics limit and the frequency evolution of the GW is unaffected. Therefore, strong lensing leads to several magnified images with the same frequency evolution, separated in time. These would appear in the data as events originating from the same sky location, with the same parameters, except for the time of arrival, the apparent luminosity distance, and an overall phase shift, which are modified as a consequences of lensing. 

\iffalse
If identified, GW lensing could open up some new exciting possibilities. For example, the detection of multiple images leads to an enhanced sky position determination. One can then correlate the GW observation with electromagnetic (ELM) observations, leading to the identification of the host galaxy for the merging black hole and enhanced cosmological measurements~\cite{Hannuksela:2020xor}. Moreover, the detection of several images means that we detect the same event multiple times, with various detector orientations. This is equivalent to multiplying the number of detectors by the number of detected images and can be used to probe the full GW polarization content and test for alternative polarizations~\cite{Goyal:2021}. 
\fi
Hence, the main idea when searching for strongly lensed GWs is to look for events that have the same intrinsic parameters and are coming from the same sky location. However, in principle one should investigate all the possible pairs one could make from the individual events, leading to an increasing list as the number of detections increases. At design sensitivity, the LIGO~\cite{TheLIGOScientific:2014jea} and Virgo~\cite{TheVirgo:2014hva} detectors could detect up to $\mathcal{O}(1000)$ events per year, leading to $~5 \times 10^5$ pairs to analyze when searching for strong lensing~\cite{Ng:2017yiu}. Currently, two parameter-estimation-based methods exist to search for lensed pairs. The first one looks for the consistency between a subset of parameters after having performed a Gaussian kernel density estimation on them~\cite{Haris:2018vmn}. This leads to the possibility of performing tests in a few minutes, but comes with a relatively large false alarm probability. The second method consists in analyzing the two data streams jointly, sampling the full joint likelihood. This leads to a high accuracy but is significantly slower~\cite{Liu:2020par,Lo:2021nae}.

 In this work, we present a new method that offers both speed and precision. We detail the Bayesian framework and show an example analysis. Our method is coded in a package called \emph{GOLUM}~\cite{Janquart:2021qov}.%, which is entirely embedded in \emph{Bilby}~\cite{Ashton:2018jfp}.

\section{Strongly lensed gravitational waves}
When a GW gets strongly lensed, it is split in multiple images, each of them being potentially observable. These images are categorized into different types, corresponding to a different overall phase shift. The modification of the lensed waveform in the frequency domain takes the form~\cite{Dai:2017huk}
\begin{equation}\label{eq:lensedWF}
    \tilde{h}_{L}^j(f; \boldtheta, \mu_j, t_{j}, n_j) = \sqrt{\mu_{j}} e^{(2\pi i f  t_{j} - i\pi n_{j} \text{sign}(f))}  \tilde{h}_{U}(f; \boldtheta) \, ,
\end{equation}
where $\tilde{h}_{U}(f; \boldtheta)$ is the unlensed waveform. Assuming GWs from coalescing binary black holes (BBHs), $\boldtheta$ is the usual set of  parameters associated with such an event, while $\mu_{j}$ is the magnification of the $j^{th}$ image and $t_j$ its arrival time. The discrete parameter $n_j$, called the Morse factor, encodes the different image types. If $n_j = 0$, the image is of type I, corresponding to a minimum of the Fermat potential; in this case the overall shape of the wave is unaffected. If $n_j = 0.5$, we have a type II image, corresponding to a saddle point of the potential, in which case the GW is Hilbert-transformed. Finally, if $n_j = 1$, we have a type III image, corresponding to a maximum of the Fermat potential; here the wave is inverted~\cite{Dai:2017huk}. The values for the time delays and magnifications depend on the nature of the lens. For a galaxy lens, the images are separated from minutes to months~\cite{Ng:2017yiu,Haris:2018vmn}, while for galaxy cluster lenses, typical time delays can be  months~\cite{Smith:2017mqu}. The time delay can be absorbed into an observed coalescence
time $t_c^{obs, j} = t_c + t_j$, and the magnification into an observed luminosity 
distance $d^{obs, j}_L = d_L/\sqrt{\mu_j}$. However, 
considering higher-order modes in the signal, the Morse factor $n_j$ can not just be absorbed into an observed coalescence phase~\cite{Ezquiaga:2020gdt,Janquart_2021}.

\section{Bayesian framework}\footnote{A more detailed description can be found in Ref.~\cite{Janquart:2021qov}.}
Under the hypothesis $\mathcal{H}_{I}$, where $I = U$ if unlensed and $I = L$ if lensed, the 
interferometer data for a pair of events are, in terms of noise and signals,
\begin{equation}\label{eq:datastream}
    d_j(t) = n_j(t) + h_I(t; \vartheta_{I,j}) \, ,
\end{equation}
where for the unlensed hypothesis $\vartheta_{U,j} = \boldtheta_j$, the usual BBH parameters, which can take different values for the two events. For the lensed hypothesis the parameters of the first event are $\vartheta_{L,1} = \mathbf{\Theta} = \{ \boldtheta_1, n_1 \}$, and defining 
\emph{relative} lensing parameters $\mathbf{\Phi} = \{\mu_{rel}, t_{21}, n_{21}\}$ with $\mu_{rel} = \mu_2/\mu_1$, $t_{21} = t_2 - t_1$, and $n_{21} = n_2 - n_1$, for the second event one has $\vartheta_{L,2} = \{ \mathbf{\Theta}, \mathbf{\Phi} \}$.

For each pair of events, under the lensing hypothesis one can compute a joint evidence 
\begin{equation}\label{eq:evidencec}
    p(d_1, d_2 | \mathcal{H}_L) = \int p(d_1 | \mathbf{\Theta}) p(d_2 | \mathbf{\Theta}, \mathbf{\Phi}) p(\mathbf{\Theta}) p(\mathbf{\Phi}) \,d\mathbf{\Theta}\,d\mathbf{\Phi} ,
\end{equation}
where $p(d_1 | \mathbf{\Theta})$, $p(d_2 | \mathbf{\Theta}, \mathbf{\Phi})$ are the likelihoods for the two images, and 
$p(\mathbf{\Theta})$, $p(\mathbf{\Phi})$ are priors. With this we can compare the two hypotheses through the \emph{coherence ratio}
\begin{equation}\label{eq:Clu}
    \mathcal{C}_U^L = \frac{p(d_1, d_2 | \mathcal{H}_L)}{p(d_1, d_2 | \mathcal{H}_U)} = \frac{p(d_2 | d_1, \mathcal{H}_L)}{p(d_2 | \mathcal{H}_U)}\frac{p(d_1 | \mathcal{H}_L)}{p(d_1 | \mathcal{H}_U)}\, ,
\end{equation}
where we used that $p(d_1, d_2 | \mathcal{H}_L) = p(d_2 |d_1,  \mathcal{H}_L) p(d_1 | \mathcal{H}_L)$ for the lensed hypothesis, and independence between the two data streams for the unlensed hypothesis.

The key point of \emph{GOLUM} is now that one can approximately identify 
\begin{equation}
    p(d_2 | d_1, \mathcal{H}_L) = \int L(\mathbf{\Phi}) p(\mathbf{\Phi})d\mathbf{\Phi} \, ,
\end{equation}
where the \emph{conditioned likelihood} $L(\mathbf{\Phi}) =  \left\langle p(d_2|\mathbf{\Theta}, \mathbf{\Phi}) \right\rangle_{p(\mathbf{\Theta}|d_{1})}$ is the likelihood of the second event averaged over the posterior samples of the first one. 
Recasting the problem in this way is what enables a significant computational speed-up of the analysis problem. 

Using the conditioned likelihood, one can evaluate the ratio of evidences  $\mathcal{C}_U^L$. However, we also want to make use of the improvement in parameter determination offered by lensing\cite{Hannuksela:2020xor}. The conditioned evidence 
gives access to the posterior distribution of the lensing 
parameters $p(\mathbf{\Phi}|d_1, d_2)$, and by reweighing,
\begin{equation}\label{eq:JointReweight}
p(\mathbf{\Theta}, \mathbf{\Phi} | d_{1}, d_{2}) \propto \frac{p(d_{2} | \mathbf{\Theta}, \mathbf{\Phi})}{p(d_{1}, d_{2} | \mathbf{\Phi})}p(\mathbf{\Theta}|d_{1})p(\mathbf{\Phi} | d_{1}, d_{2}) \, .
\end{equation}
This enables us to recover joint parameters similar to those obtained by joint parameter estimation frameworks. 
Once the first image has been analyzed, the second image analysis and the reweighing process take $\mathcal{O}(30)$ CPU minutes, hence the computational time is essentially that of a single event analysis. In addition, these calculations are not restricted to pairs and can be extended to any number of images~\cite{Janquart:2021qov}.

\section{Example analysis for a quadruply lensed gravitational wave event}\label{sec:ExampleAnalysis}
As an illustration, we inject a signal from a precessing BBH merger generated with \textsc{IMRPhenomPv2}~\cite{Khan:2015jqa} into synthetic stationary Gaussian noise for a network of detectors made of the two LIGO detectors and Virgo, at design sensitivity~\cite{aLIGOdesign,TheVirgo:2014hva}. The event has masses of $36 \, M_{\odot}$ and $29.2\, M_{\odot}$ and spin magnitudes 0.4 and 0.3, with tilt angles of 0.5 rad and 1 rad, respectively. The spin vectors' azimuthal angle is 1.7 rad, while the precession angle about the orbital angular momentum is 0.3 rad. The apparent luminosity distance for the first image is $1500$ Mpc. The inclination angle is 0.4 rad, the polarization angle is 2.66 rad, and the unlensed coalescence phase is 1.3 rad. For the sky position of the event, the right ascension is 1.375 rad, and the declination is $-1.21$ rad. The first image is a type II image, with a Morse factor of 0.5. For the first image analysis, we use a uniform prior in chirp mass and mass ratio, and time of arrival. The prior on luminosity distance is uniform in comoving volume. All the other priors are standard priors for a BBH parameter estimation analysis. 

We generate three additional lensed images and link them via the relative lensing parameters. The second image has $\mu_{rel} = 2$, $t_{21} = 14\, \rm{hr}$, $n_{21} = 0.5$, the third image has  $\mu_{rel} = 4$, $t_{31} = 16\, \rm{hr}$, $n_{31} = 0$, and the fourth  $\mu_{rel} = 5$, $t_{41} = 21\, \rm{hr}$, $n_{41} = 1$. For each of these images, we perform a nested sampling run to get the unlensed evidence. For the lensed hypothesis, we first perform the analysis of the first image using the priors specified above, and we use the posterior samples in a conditioned likelihood for the second image analysis. For this image, and all the others, the prior for the relative magnification is uniform between $0$ and $10$, the prior for the time delay is uniform in $[t_{j1} - 0.1, t_{j1} + 0.1]$, and the prior on the Morse factor difference is discrete uniform in $\{0, 0.5, 1, 1.5\}$. After its analysis, we get the posterior for the second image. Reweighed samples from the image 1 and 2 analyses can then be used in a conditioned likelihood to analyze the third image, and similarly for the fourth image. The process is thus made of a succession of \emph{GOLUM} runs and reweighing, one for each additional image, reducing significantly the cost of multiple image analyses.

\begin{figure}[t]
\centering
    \includegraphics[keepaspectratio, width=0.5\textwidth]{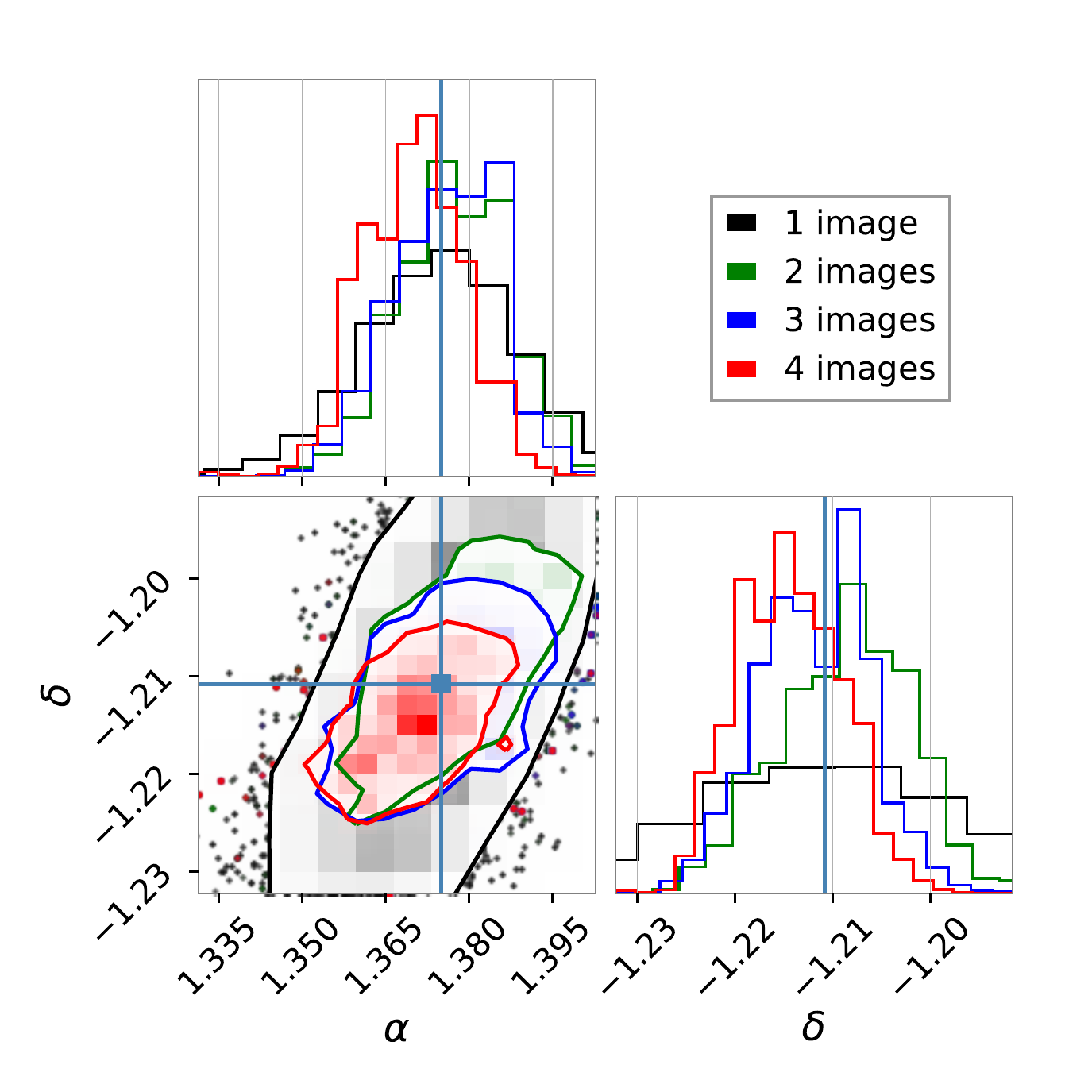}
    \caption{The $90\%$ credible region of the sky location of a strongly lensed gravitational-wave singlet (black), pair (green), triplet (blue),  and quadruplet (red). 
    There is a clear improvement in the sky localization with the addition of every gravitational-wave image. The final $90\%$ confidence sky area is $\sim 2\deg^{2}$ in this example, down from $\sim 20\deg^2$ for the singlet. An improved sky localization would be particularly useful for lensed host galaxy localization\protect\cite{Hannuksela:2020xor} .}
    \label{fig:skyLoc4imgs}
\end{figure}

To illustrate the benefit of lensing on the posterior distribution, we focus on the evolution of the sky localization as a function of the number of images. The evolution of these posteriors can be seen in Fig.~\ref{fig:skyLoc4imgs}, where with each additional image the sky location area is reduced, going from $\sim 20 \, \rm{deg}^2$ for one image to $\sim 2 \, \rm{deg}^2$ for four images at 90\% confidence. This is in agreement with the expected improvement in sky localization for lensed BBHs when multiple images are detected~\cite{Hannuksela:2020xor}.

\section{Conclusions}
We have introduced \emph{GOLUM}, a fast and precise methodology to analyze strongly lensed gravitational waves, and have shown its use in an example analysis. The method enables us to have fast analyses with good accuracy, helping to tackle the issue of an increasing number of event pairs to be analyzed. In addition, it enables the analysis of more than two images. These benefits also make the methodology useful for science case studies related to strong lensing, where one wants to inject a large number of signals~\cite{Janquart_2021}.

\section*{Acknowledgments}
J.J, K.H, and C.V.D.B are supported by the research program of the Netherlands Organisation for Scientific Research (NWO). O.A.H was partially supported by grants from the Research Grants Council of the Hong Kong (Project No. CUHK 14306218), The Croucher Foundation of Hong Kong and Research Committee of the Chinese University of Hong Kong.
The authors are grateful for computational resources provided by the LIGO Laboratory and supported by the National Science Foundation Grants No. PHY-0757058 and No. PHY-0823459.

\end{document}